\documentclass[10pt]{article}
\usepackage{graphicx}
\usepackage{epstopdf}
\usepackage{amsmath}%
\DeclareSymbolFont{bbold}{U}{bbold}{m}{n}
\DeclareSymbolFontAlphabet{\mathbbold}{bbold}

\usepackage{graphicx}
\usepackage{float}

\usepackage{amsfonts}%
\usepackage{amssymb}%
\usepackage{eucal}
\usepackage{youngtab}
\usepackage{hyperref}
\usepackage{makeidx}
\makeindex
\usepackage{nomencl}
\makenomenclature
\usepackage{diagbox}
\usepackage{theorem}
\usepackage{url}
\usepackage{array}

\usepackage{color}
\usepackage{upgreek}
\usepackage{epsfig}
\usepackage{graphicx}
\usepackage{verbatim}
\usepackage{float}
\usepackage{braket}
\usepackage{appendix}
\usepackage[margin=2.5cm]{geometry}
\usepackage{parskip}
\usepackage[titles]{tocloft}
\usepackage{pst-node}
\usepackage{amsmath}

\newcommand{\qed}{\nobreak \ifvmode \relax \else
      \ifdim\lastskip<1.5em \hskip-\lastskip
      \hskip1.5em plus0em minus0.5em \fi \nobreak
      \vrule height0.75em width0.5em depth0.25em\fi}
\def\VZ#1{\phantom{-}\ifx\relax#1\relax0\else\rnode{#1}{0}\fi}
\usepackage{blkarray}
\usepackage{enumerate}

\begin{document}
\rightline{LTH-1079}
\rightline{March 2016}
	
\begin{center}
\LARGE
The Flipped $SU(5)$ String Vacua Classification: \\ A Variation Of The $SO(10)$ Breaking Basis Vector\end{center}
\begin{center}
\end{center}
\begin{center}
\Large
Hasan Sonmez\footnote{Hasan.Sonmez@Liverpool.ac.uk}

\small
\it Department of Mathematical Sciences, University of Liverpool, Liverpool L69 7ZL, United Kingdom
\end{center}
\begin{abstract}
In this paper, an extension of the classification of flipped $SU(5)$ heterotic-string vacua from \cite{frs} with a variation of the $SO(10)$ breaking $\alpha$ basis vector is presented. A statistical sampling in the space of $2^{45}$ flipped $SU(5)$ vacua is explored, where $10^{11}$ distinct GGSO projection configurations are scanned in comparison to the $10^{12}$ GGSO distinct coefficients scanned in the space of $2^{44}$ vacua in \cite{frs}. A JAVA code, akin to the one used for the classification in \cite{frs}, was implemented to explore these. Results presented here indicate that no three-generation exophobic vacua exist, which was also found to be the case in \cite{frs} as all odd generations were projected out. This paper will also study the details on the comparison between the two classifications achieved and reflect on the future directions in the quest for finding three-generation exophobic flipped $SU(5)$ heterotic-string models.
\end{abstract}
\section{\emph{Introduction}}
To date vast amounts of work have been carried out using techniques in string model building to construct physical spectrums consisting of three generations. Concrete examples of analyzing semirealistic string vacua can be found in both symmetric and asymmetric ${\mathbb{Z}}_2 \times {\mathbb{Z}}_2$ orbifold compactifications, especially using the four-dimensional free-fermionic construction \cite{ABKKLT1987a,ABKKLT1987b,KLT,PhDThesis} given by a worldsheet approach in the heterotic-string setting. Initially, the asymmetric ${\mathbb{Z}}_2 \times {\mathbb{Z}}_2$ free-fermionic orbifold models corresponding to the ${\cal N}=(2,0)$ super-conformal worldsheet symmetry were studied during the late 1980s. The gauge group was the observable $E_8$ in these cases, which was then broken to the $SO(10)$ symmetry followed by another breaking to its subgroup in the NAHE-based\footnote{Nanopoulos, Antoniadis, Hagelin and Ellis} structure \cite{NAHE1993}. These subgroups consisted of the flipped $SU(5)$ gauge group \cite{RevampAEHN,FSU51993}, $SU(3)\times SU(2)\times U(1)^2$ standard-like gauge group \cite{SLMa,SLMb,SLMc,SLMd,SLMe}, $SU(3)\times SU(2)^2\times U(1)$ left-right symmetric gauge group \cite{LRS2001a,LRS2001b,NonSusyPaper}, and the $SO(6)\times SO(4)$ Pati-Salam gauge group \cite{ALR1990a,ALR1990b}. This aided in the development of the free-fermionic model building contemporary research on exploring large classes of string vacua. In the late 1990s, work focused on the symmetric ${\mathbb{Z}}_2 \times {\mathbb{Z}}_2$ free-fermionic orbifolds. These helped derive tools for the classifications, such as during the last decade for the type II superstrings \cite{GKR1999} and then extended to the heterotic-string free-fermionic construction \cite{FKNR2004a,FKNR2004b,FKR2007a,FKR2007b,FKR2007c}.

Untill now the classifications of the free-fermionic vacua have extracted many viable phenomenological models at the string scale. Initially, in the study of $E_6$ and $SO(10)$ GUT symmetries, the existence of a symmetry called the spinor-vector duality was shown in the space of ${\mathbb{Z}}_2$ and ${\mathbb{Z}}_2 \times {\mathbb{Z}}_2$ string models. This symmetry generated the exchange of the vectorial $\bf{10}$ representations with the representations of the spinorial $\bf{16}$ plus anti-spinorial $\bf{\overline{16}}$ under the $SO(10)$ gauge group \cite{FKR2007a,FKR2007b,FKR2007c,XMAP1993a,XMAP1993b,spinvecduala,spinvecdualb,spinvecdualc,spinvecduald,Rizos2011}. Such findings motivated further classifications involving $SO(10)$ subgroups, where the Pati-Salam models were investigated \cite{ACFKR2011a,ACFKR2011b}. Here the sectors containing only the massless states revealed the existence of exophobic string vacua and therefore leaving the fractionally charged fermions (exotics) \cite{frs,PhDThesis,WW1985a,WW1985b,Schellekens1990,FC1992a,FC1992b,FC1992c,Halyo2000} to exist in the massive spectrum. It was also shown to generate an abundance of three-generation models. A detailed example of a phenomenologically viable three-generation Pati-Salam model was studied in Ref. \cite{CFR2011}. Another detailed model was then studied in Ref. \cite{SU6SU2}, which enhanced the Pati-Salam models to the $SU(6)\times SU(2)$ \cite{SU6SU2} maximal subgroup of the $E_6$ symmetry. It was also observed that an exophobic three-generation model existed, admitting an anomaly-free family universal $U(1)$ symmetry, which was additional to the $U(1)$ generators of the $SO(10)$ GUT symmetry \cite{AnomalyFreeU1a,AnomalyFreeU1b,AnomalyFreeU1c,AnomalyFreeU1d}. Later, the classifications for the flipped $SU(5)$ \cite{frs,PhDThesis} and the $SU(4) \times SU(2) \times U(1)$ \cite{PhDThesis,SU421} $SO(10)$ subgroup models were carried out where it was shown to contain no exophobic three-generation string vacua. In the flipped $SU(5)$ models only exophobic even generations existed, as all the exophobic odd generations where projected out. As for finding three-generation flipped $SU(5)$ vacua, models needed to be exophilic. The $SU(4) \times SU(2) \times U(1)$ models on the other hand, did not consist of any exophobic three-generation vacua, since in the massless spectrum all the right-handed particles were projected out, and therefore it was not possible to find any generations. In fact, in the NAHE-based basis vectors in Ref. \cite{NAHESU421}, it was similarly shown that three-generation models are forbidden for the $SU(4) \times SU(2) \times U(1)$ gauge group. For a more detailed analysis on the free-fermionic landscape, see for example Refs. \cite{PhDThesis} and \cite{DiscretePaper}.

The free-fermionic classifications carried out in Refs.  \cite{frs,PhDThesis,GKR1999,FKNR2004a,FKNR2004b,FKR2007a,FKR2007b,FKR2007c,Rizos2011,ACFKR2011b,SU421} have developed the methodologies for providing a pivotal tool set, leading to an elegant way to analyze large sets of string vacua specific to the free-fermionic construction. The analysis of large sets of string vacua carried out by other groups can be found in Ref. \cite{StatStudy1989}. In this paper, these techniques are used to analyze a larger class of string vacua belonging to the new flipped $SU(5)$ models. Here, the $SO(10)$ symmetry is constructed as in Ref. \cite{frs} that is then broken to a different flipped $SU(5)$ gauge group given by a variation of the $SO(10)$ symmetry breaking $\alpha$ basis vector. The difference between the flipped $SU(5)$ models given here and those in Ref. \cite{frs}, is that the hidden gauge groups are broken differently. After discussing the classification methodology, detailed properties of the classification of these new flipped $SU(5)$ models are given, followed by the discussion of the results for the existence of three-generation models in the free-fermionic landscape. A comparison is then drawn between the two explaining future directions to be taken.

\section{\emph{The Free-Fermionic Flipped $SU(5)$ Construction}}
The four-dimensional free-fermionic construction is represented in the light-cone gauge; this consists of $20$ real left-moving worldsheet fermions $\psi^\mu_{1,2}, \chi^{1,\dots,6},y^{1,\dots,6}, \omega^{1,\dots,6}$ and $44$ real right-moving real worldsheet fermions $\overline{y}^{1,\dots,6},\overline{\omega}^{1,\dots,6}$,
$\psi^{1,\dots,5}$, $\overline{\eta}^{1,2,3}$, $\overline{\phi}^{1,\dots,8}$. The worldsheet of the free-fermionic construction is given by a torus, where as the worldsheet fermions in the vacuum to vacuum amplitude are parallel transported along the two noncontractible loops, they acquire a phase. Therefore, under the modular invariance constraints \cite{ABKKLT1987a,ABKKLT1987b,PhDThesis} imposed, the construction allows for model building  with specifically chosen gauge groups. This is achieved by assigning particular choices of boundary conditions to the fermions in each set of basis vectors. The basis vectors themselves, form an additive group; this then generates the entire string spectrum consisting of $2^{N}$ sectors in the space of $\Xi$, where $N$ is the number of basis vectors. Here a sector is defined by a specific linear combination of the basis vectors and the total space $\Xi$ is given by a set of all possible linear combinations. Additionally, the basis vectors also induce a specific generalized GSO projection on any given string state $|S_{\xi}>$, performed by an action. This can be written as
\begin{equation*}\label{gso}
	e^{i\pi v_i\cdot F_{\xi}} |S_{\xi}> = \delta_{{\xi}}\ C \binom {\xi} {v_i}^* |S_{\xi}>,
\end{equation*}
where the space-time spin statistics is given by $\delta_{{\xi}}=\pm1$ and the fermion number operator is given by $F_{\xi}$. The GGSO projection coefficients $ C \binom {\xi}{v_i}=\pm1,\pm i$ in the equation above, are used to produce distinct string models by varying the GGSO projection coefficients. In summary, a model from the free-fermionic construction is given by a set of basis vectors $v_{1},\dots,v_{N}$, a set of $2^{N(N-1)/2}$ string vacua and independent $C \binom{v_i}{v_j}, i>j$ GGSO projection coefficients.

\subsection{\emph{The $\alpha$ Basis Vector}}
As highlighted in Ref. \cite{frs}, the choice of the basis vectors that breaks the
$SO(10)$ symmetry to the $SU(5)\times U(1)$ gauge group is not unique. Although, the breaking to the $SU(5)\times U(1)$ symmetry is unique, the difference lies in the breaking of the hidden gauge group. Here, the boundary condition assignments of the three complex worldsheet fermions $\eta^{1,2,3}= 1/2$ are constrained by the term $b_j \cdot\alpha  = 0\mod 1$ given by modular invariance, whereas the boundary condition assignments of the five complex worldsheet fermions $\psi^{1,...,5}= 1/2$ are fixed to construct the $SU(5)\times U(1)$ gauge symmetry. Therefore, the variation of the $\alpha$ basis vectors exists in the assignment of the boundary conditions of the remaining eight worldsheet complex fermions $\phi^{1,...,8}$. In this case, modular invariance restricts the assignment of $1/2$ boundary conditions to zero, four or eight worldsheet complex fermions of $\phi^{1,...,8}$. This leads to breaking the hidden gauge group differently. The case with zero enhances the $SU(5)\times U(1)$ gauge symmetry to $SO(10)$ and therefore is of no importance, the case with four was studied in Ref. \cite{frs} and in this paper, the $\alpha$ basis vector assigned with $1/2$ boundary conditions to the eight $\phi^{1,...,8}$ complex worldsheet fermions is presented, which is given by

$$\alpha \, = \{ \overline{\eta}^{1,2,3} = \frac{1}{2}, \overline{\psi}^{1,...,5} = \frac{1}{2}, \overline{\phi}^{1,...,8} = \frac{1}{2}\}.$$

\subsection{\emph{The Flipped $SU(5)$ Basis Vectors}}
The $SU(5)\times U(1)$ models in the free-fermionic construction are obtained by the breaking of the $SO(44)$ gauge group to the $SO(10)$ gauge group and then to the flipped $SU(5)$ by a series of basis vectors given below:
\begin{eqnarray}
v_1=S&=&\{\psi^\mu,\chi^{1,\dots,6}\},\nonumber\\
v_{1+i}=e_i&=&\{y^{i},\omega^{i}|\overline{y}^i,\overline{\omega}^i\}, \
i=1,\dots,6,\nonumber\\
v_{8}=b_1&=&\{\chi^{34},\chi^{56},y^{34},y^{56}|\bar{y}^{34},
\overline{y}^{56},\overline{\eta}^1,\overline{\psi}^{1,\dots,5}\},\label{basis}\\
v_{9}=b_2&=&\{\chi^{12},\chi^{56},y^{12},y^{56}|\overline{y}^{12},
\overline{y}^{56},\overline{\eta}^2,\overline{\psi}^{1,\dots,5}\},\nonumber\\
v_{10}=z_1&=&\{\overline{\phi}^{1,\dots,4}\},\nonumber\\
v_{11}=z_2&=&\{\overline{\phi}^{5,\dots,8}\},\nonumber\\
v_{12} \,\,\, =\alpha \, &=& \{ \overline{\eta}^{1,2,3} = \frac{1}{2},\overline{\psi}^{1,...,5} = \frac{1}{2},\overline{\phi}^{1,...,8} = \frac{1}{2}\}. \nonumber
\end{eqnarray}
Here, the $SO(44)$ gauge symmetry is generated with the basis vector $S$ that also makes the models ${N} = 4$ supersymmetric. In order to incorporate all the possible symmetric shifts from the six internal
coordinates, the next six vectors $e_{i}$ for $i=1,...,6$ are formed, where the ${N = 4}$ supersymmetry is left intact together with the breaking of the $SO(44)$ gauge group to the $SO(32) \times U(1)^6$ symmetry. The set $\{b_1, b_2\}$ leads to the breaking of ${N} = 4$ to ${N} = 1$ supersymmetry; these vectors correspond to the ${\mathbb{Z}}_2 \times {\mathbb{Z}}_2$ orbifold twists that also break the $U(1)^6$ together with decomposing the $SO(32)$ gauge symmetry to $SO(10) \times U(1)^3 \times SO(16)$. The $SO(10) \times U(1)^3$ group constructed here is defined to represent the observable symmetry and the other $SO(16)$ group is defined to represent the hidden symmetry. The set $\{z_1, z_2\}$ is formed to further break the $SO(16)$ gauge group to the $SO(8)_{1}\times SO(8)_{2}$ symmetry. Last, as discussed before the $\alpha$ basis vector breaks the $SO(10)$ symmetry to the flipped $SU(5)$ group and then the overall gauge group generated by the gauge bosons in the untwisted Neveu Schwarz sector is given by
$$
\underbrace{SU(5)\times U(1)\times{U(1)}^3}_{Observable}\times \underbrace{SU(4)_{1} \times U(1)_{1} \times SU(4)_{2} \times U(1)_{2}}_{Hidden}
$$
Additionally, in order to satisfy the ABK\footnote{Antoniadis, Bachas and Kounnas} rules \cite{ABKKLT1987a,ABKKLT1987b,KLT,PhDThesis}, the basis vector $\mathbbold{1}$ must consist in the additive group given by the above basis vectors. This is produced by the following sum
\begin{equation*} \label{BasisOne}
\mathbbold{1} = S+\sum_{i=1}^{6}e_{i}+2\alpha.
\end{equation*}

\section{\emph{Classification Methodology}}\label{classification}
The scanning of vast amounts of free-fermionic string vacua is facilitated by an advanced classification methodology, where the twisted sectors are identified and then split into the observable, hidden and exotic sectors. These sectors are computed by the use of a JAVA code which checks for viable phenomenological properties. The observable sectors are scanned for chiral matter and Higgs states, whereas the hidden sectors only contain $SU(5)$ singlets and therefore are all neutral under all the observable $U(1)$'s. However, the exotic sectors all contain fractionally charged massless fermion states \cite{frs,PhDThesis,WW1985a,WW1985b,Schellekens1990,FC1992a,FC1992b,FC1992c} under the $SU(5) \times U(1)$ symmetry; as these representations do not fall into the Standard Model \cite{Halyo2000}, they are all avoided in the scan for viable free-fermionic string models.

\subsection{\emph{GGSO Projections Coefficients}}
The free-fermionic string vacua are primarily defined by the GGSO projection coefficients $c\binom{v_i}{v_j}$ contained in the one-loop partition function; these are all given as free parameters. In order to extract viable phenomenological models some of these GGSO projection coefficients are fixed. In total taking the coefficients as spanning a $12\times12$ matrix and then imposing modular invariance, only the elements belonging to $i \geq j$ are independent. Moreover, the 66 lower triangle elements are defined by the corresponding 66 upper triangle elements. Including the remaining 12 diagonal elements, only 78 independent coefficients are left as free parameters. Further requirements such as imposing ${N}=1$ space-time supersymmetry, fix the following 11 coefficients:
\begin{eqnarray*}
	\label{Susyfixcoefficients}
	C\binom{S}{S} = C\binom{S}{e_{i}} = C\binom{S}{b_{k}} =
	C\binom{S}{z_k} = C\binom{S}{\alpha} = -1 , &&\\
	i=1,...,5, \, k = 1,2.\nonumber
\end{eqnarray*}
This reduces the amount of free parameters to 67. The diagonal terms are also fixed by modular invariance and are given as
\begin{align*}
	1 &= \prod_{\substack{j=1 \\ i \neq j}}^{6}
	C \binom {e_i} {e_j}, \nonumber\\
	\label{constraints}
	C \binom {b_k} {b_k} &= - \prod_{i=1}^{6} C \binom {b_k} {e_i}, \ \ \ k=1,2,\\
	C \binom {z_k} {z_k} &= - \prod_{i=1}^{6} C \binom {z_k} {e_i}, \ \ \ k=1,2,\nonumber\\
	C \binom {\alpha} {\alpha} &= - \prod_{i=1}^{6}
	C \binom {\alpha} {e_i}.\nonumber
\end{align*}
Therefore, this leads to another 11 coefficients being fixed. In fact, during the developments of the free-fermionic classifications, it was found that additionally coefficients do not affect the free-fermionic string spectra. Specific to the models studied here, these are the following 11 coefficients:
\begin{eqnarray*}
	\label{freecoefficients}
	C\binom{e_i}{e_i}, C\binom{e_3}{b_1}, C\binom{e_4}{b_1},
	C\binom{e_1}{b_2}, C\binom{e_2}{b_2}, C\binom{b_1}{b_2}, &&\\
	i=1,...,6.
	\nonumber
\end{eqnarray*}
As these given coefficients are ineffective on the string spectrum, they are fixed arbitrarily, which then leaves in total 45 free parameters defining the full spectrum. This corresponds to $2^{45} \approx 3.52 \times 10^{13}$ vacua in this class of free-fermionic superstring models.

\subsection{\emph{Observable Matter Spectrum}}
The matter content arising from the string spectrum in the free-fermionic models, is from the $\bf{27}$ representation belonging to the $E_6$ symmetry that is then broken to the $SO(10)$ GUT symmetry at the string scale. This leads to the particle content of the Standard Model to be found in the $\bf{16}$ spinorial representations, whereas the light Higgs states are found in the $\bf{10}$ vectorial representations. The specific sectors containing the chiral matter content in the symmetric free-fermionic constructions are generated from the following equations:
\begin{eqnarray} \label{obspin}
	B_{pqrs}^{(1)}&=& S + {b_1 + p e_3+ q e_4 + r e_5 + s e_6} \nonumber\\
	&=&\{\psi^\mu,\chi^{12},(1-p)y^{3}\overline{y}^3,p\omega^{3}\overline{\omega}^3,
	(1-q)y^{4}\overline{y}^4,q\omega^{4}\overline{\omega}^4, \nonumber\\
	& & ~~~(1-r)y^{5}\overline{y}^5,r\omega^{5}\overline{\omega}^5,
	(1-s)y^{6}\overline{y}^6,s\omega^{6}\overline{\omega}^6,
	\overline{\eta}^1,\overline{\psi}^{1,..,5}\},
	\\
	B_{pqrs}^{(2)}&=& S + {b_2 + p e_1+ q e_2 + r e_5 + s e_6},
	\label{twochiralspinorials}
	\nonumber\\
	B_{pqrs}^{(3)}&=& S + {b_3 + p e_1+ q e_2 + r e_3 + s e_4}. \nonumber
\end{eqnarray}
Here $p,q,r,s=0,1$ and $b_3=b_1+b_2+2\alpha+z_1+z_2$. These equations consist of 48 sectors that either produce the left- and right-handed particles from the $\bf{16}$ or the $\overline{\bf{16}}$ spinorial representation of the $SO(10)$. For the flipped $SU(5)$ models studied in this paper, the $SO(10)$ $\bf{16}$ and $\overline{\bf{16}}$ representations decompose under the $SU(5) \times U(1)$ symmetry as follows:
\begin{eqnarray*}
	\textbf{16} &= &\left(\overline{\textbf{5}},
	-{\tfrac{{3}}{{2}}}\right) + 
	\left(\textbf{10},+{\tfrac{{1}}{{2}}}\right) + 
	\left(\textbf{1},+{\tfrac{{5}}{{2}}}\right), \nonumber \\
	\overline{\textbf{16}} &= &\left(\textbf{5},
	+{\tfrac{{3}}{{2}}}\right) + 
	\left(\overline{\textbf{10}},-{\tfrac{1}{2}}\right)
	+ \left(\textbf{1},-{\tfrac{{5}}{{2}}}\right).
\end{eqnarray*}
In order to extract the Standard Model particles from the above $SU(5)$ representations, the weak hypercharge and the electromagnetic charge are defined from the following normalizations:
\begin{eqnarray*}
	Y &=& \frac{1}{3} (Q_1 + Q_2 + Q_3) + \frac{1}{2} (Q_4 + Q_5), \nonumber\\
	Q_{em} &=& Y + \frac{1}{2} (Q_4 - Q_5),
\end{eqnarray*}
where $Q_{i}=\pm \frac{1}{2}$ are the charges arising from the $\overline{\psi}^{i}$ complex fermions for $i = 1,...,5$. Furthermore, the matter content in the free-fermionic flipped $SU(5)$ models are summarized in the following table:
\begin{center}
	\begin{tabular}{|c|c|c|c|c|}
		\hline
		Representation & $\overline{\psi}^{1,2,3}$ &
		$\overline{\psi}^{4,5}$ & $Y$ & $Q_{em}$ \\
		\hline \hline
		& $(+,+,+)$ & ($+,-$)& 1/2& 1,0\\
		$\left( \, \textbf{5} \, , \, +\frac{{3}}{{2}} \, \right)$ & ($+,+,-$)& $(+,+)$ & 2/3& 2/3\\ 
		\hline
		& ($+,-,-$)& $(-,-)$ & -2/3& -2/3\\
		$\left( \, \overline{\textbf{5}} \, , \, -\frac{{3}}{{2}} \, \right)$ & $(-,-,-)$ & ($+,-$)& -1/2& -1,0\\ 
		\hline
		& $(+,+,+)$ & $(-,-)$ & 0& 0\\
		$\left(\textbf{10},+\frac{{1}}{{2}}\right)$& ($+,-,-$)& $(+,+)$ & 1/3& 1/3\\
		& ($+,+,-$)& ($+,-$)& 1/6& -1/3,2/3\\ 
		\hline
		& ($+,+,-$)&
		$(-,-)$ & -1/3& -1/3\\
		$\left(\overline{\textbf{10}},-\frac{{1}}{{2}}\right)$ & ($+,-,-$)& ($+,-$)& -1/6& 1/3,-2/3\\
		& $(-,-,-)$ & $(+,+)$ & 0& 0\\ 
		\hline
		$( \, \textbf{1} \, , \, +\frac{{5}}{{2}} \, )$&
		$(+,+,+)$ & $(+,+)$ & 1& 1\\ 
		\hline
		$( \, \textbf{1} \, , \, -\frac{{5}}{{2}} \, )$
		& $(-,-,-)$ & $(-,-)$ & -1& -1\\ 
		\hline
	\end{tabular}
\end{center}
From this table, it is now shown that the Standard Model particle representations are decomposed to the $SU(3) \times SU(2) \times U(1)$ gauge group from the flipped $SU(5)$ symmetry as follows:
\begin{align*}
	\left( \,  \,  \overline{\textbf{5}} \,  ,-\frac{3}{2}\right)&
	=\left(\overline{\textbf{3}},\textbf{1},-\frac{2}{3}\right)_{u^c}+\left(\textbf{1},\textbf{2},-\frac{1}{2}\right)_{L}, \nonumber \\
	\left(\textbf{10},+\frac{1}{2}\right)&=\left(\textbf{3},\textbf{2},+\frac{1}{6}\right)_{Q} \, 
	+\left(\overline{\textbf{3}},\textbf{1},+\frac{1}{3}\right)_{d^c}+\left(\textbf{1},\textbf{1},0\right)_{\nu^c},\\
	\left( \,  \,  \textbf{1} \, ,+\frac{5}{2}\right)&=\left(\textbf{1},\textbf{1},+1 \, \right)_{e^c}, \nonumber
\end{align*}
where the subscripts are the standard notations defining the left-handed quark and lepton doublets as $Q$ and $L$ respectively and the right-handed quark and lepton singlets as $d^c,~u^c,~e^c$ and $\nu^c$ respectively.

\subsection{\emph{The Enhancement Sectors}}
In the free-fermionic construction space-time vector bosons arise from the untwisted NS sector,  which generates the $SO(44)$ gauge group. This group is then broken to the flipped $SU(5)$ symmetry with a series of GGSO projections with respect to the basis vectors as defined in Eq. (\ref{basis}). However, the flipped $SU(5)$ gauge group can then be enhanced to another gauge group depending on the GGSO projection coefficients chosen. This property emerges due to the extra space-time vector bosons arising from the following sectors:

\begin{equation}\label{ggsectors}
	\mathbf{G} =
	\left\{ \begin{array}{ccccc}
		\,\,\,\, z_1          ,&
		\,\,\,\, z_2          ,&
		\,\,\,\, z_1 + z_2    ,&
		\,\,\,\, z_1+z_2 + 2\alpha,\\
		\,\,\,\,\, \alpha       ,&
		\,\,\,\,\, z_1 + \alpha ,&
		\,\,\, z_2 + \alpha ,&
		\,\, z_1 + z_2 + \alpha,\\
		\,\, 3\alpha       ,&
		\,\,\,\,\,\,\,\, z_1 + 3\alpha ,&
		\,\,\,\,\,\, z_2 + 3\alpha ,&
		\,\,\, z_1 + z_2 + 3\alpha
	\end{array} \right\}. \nonumber
\end{equation}

In order to preserve the flipped $SU(5)$ symmetry, the imposed restriction in this paper, is that the only gauge bosons that remain in the spectrum are those that are obtained from the untwisted NS sector and the sectors that only enhance the hidden gauge group. The reason for allowing the enhancement of the hidden gauge group is to provide extra freedom to find three-generation exophobic models which were not found in Ref. \cite{frs} as will be discussed later.

\subsection{\bf{Projectors}}
In Eq. (\ref{obspin}), the sectors $B_{pqrs}^{(1,2,3)}$ produce matter states subject to surviving the GGSO projection. To decide if the state is projected in or out, the GGSO projection coefficients in relation to the basis vectors $e_1$, $e_2$, $z_1$ and $z_2$ are checked. Therefore, generic formulas called the projectors are defined, where $P=1$ is the indicator for surviving states and $P=0$ for projected out states. These projectors are given as
\begin{align*}
	P_{pqrs}^{(1)} &= \frac{1}{16} 
	\left( 1-C \binom {e_1} {B_{pqrs}^{(1)}}\right) . 
	\left( 1-C \binom {e_2} {B_{pqrs}^{(1)}}\right) . 
	\left( 1-C \binom {z_1} {B_{pqrs}^{(1)}}\right) . 
	\left( 1-C \binom {z_2} {B_{pqrs}^{(1)}}\right), \nonumber\\
	P_{pqrs}^{(2)} &= \frac{1}{16} 
	\left( 1-C \binom {e_3} {B_{pqrs}^{(2)}}\right) . 
	\left( 1-C \binom {e_4} {B_{pqrs}^{(2)}}\right) . 
	\left( 1-C \binom {z_1} {B_{pqrs}^{(2)}}\right) . 
	\left( 1-C \binom {z_2} {B_{pqrs}^{(2)}}\right),\\
	P_{pqrs}^{(3)} &= \frac{1}{16} 
	\left( 1-C \binom {e_5} {B_{pqrs}^{(3)}}\right) . 
	\left( 1-C \binom {e_6} {B_{pqrs}^{(3)}}\right) . 
	\left( 1-C \binom {z_1} {B_{pqrs}^{(3)}}\right) . 
	\left( 1-C \binom {z_2} {B_{pqrs}^{(3)}}\right). \nonumber
\end{align*}
In order to incorporate this into a JAVA code, these projectors are given in the form of systems of linear equations, where $p$, $q$, $r$ and $s$ are the unknowns. These systems of linear equations are written as
\begin{align*} \label{matrix}
	\begin{pmatrix} (e_1|e_3)&(e_1|e_4)&(e_1|e_5)&(e_1|e_6)\\
		(e_2|e_3)&(e_2|e_4)&(e_2|e_5)&(e_2|e_6)\\
		(z_1|e_3)&(z_1|e_4)&(z_1|e_5)&(z_1|e_6)\\
		(z_2|e_3)&(z_2|e_4)&(z_2|e_5)&(z_2|e_6) \end{pmatrix}
	\begin{pmatrix} p\\q\\r\\s\end{pmatrix} &=
	\begin{pmatrix} (e_1|b_1)\\
		(e_2|b_1)\\
		(z_1|b_1)\\
		(z_2|b_1)
	\end{pmatrix}, \nonumber
	\\[0.3cm]
	\begin{pmatrix} (e_3|e_1)&(e_3|e_2)&(e_3|e_5)&(e_3|e_6)\\
		(e_4|e_1)&(e_4|e_2)&(e_4|e_5)&(e_4|e_6)\\
		(z_1|e_1)&(z_1|e_2)&(z_1|e_5)&(z_1|e_6)\\
		(z_2|e_1)&(z_2|e_2)&(z_2|e_5)&(z_2|e_6) \end{pmatrix}
	\begin{pmatrix} p\\q\\r\\s\end{pmatrix} &=
	\begin{pmatrix} (e_3|b_2)\\
		(e_4|b_2)\\
		(z_1|b_2)\\
		(z_2|b_2)
	\end{pmatrix},
	\\[0.3cm]
	\begin{pmatrix} (e_5|e_1)&(e_5|e_2)&(e_5|e_3)&(e_5|e_4)\\
		(e_6|e_1)&(e_6|e_2)&(e_6|e_3)&(e_6|e_4)\\
		(z_1|e_1)&(z_1|e_2)&(z_1|e_3)&(z_1|e_4)\\
		(z_2|e_1)&(z_2|e_2)&(z_2|e_3)&(z_2|e_4) \end{pmatrix}
	\begin{pmatrix} p\\q\\r\\s\end{pmatrix} &=
	\begin{pmatrix} (e_5|b_3)\\
		(e_6|b_3)\\
		(z_1|b_3)\\
		(z_2|b_3)
	\end{pmatrix}. \nonumber
\end{align*}
Using methods developed in linear algebra, the JAVA code using these equations can easily check the condition Rank(Matrix[$\Delta^{i}$])$=$Rank(AugmentedMatrix[$\Delta^{i},Y^{i}$]). Here when the condition is equal, the computer code would instantly realize that there are $2^{4-\mbox{Rank}(\mbox{Matrix}[\Delta^{i}])}$ solutions. If a sufficient amount of solutions were found, the code would then do more extensive checks to test if the models have any viable phenomenological properties.

\section{\emph{Classification}}
An elegant technique in the classification, is to use the algebraic expressions as discussed in the previous section to analyze the entire massless free-fermionic string spectrum. These expressions are then transformed into matrix equations that are solved by a JAVA code that randomly selects a configuration of a set of GGSO projection coefficients. The JAVA code is used to scan models in the space of $2^{45} \approx 3.52 \times 10^{13}$ string vacua. Since the entire string vacua is a large space, a random generator was used to scan $10^{11}$ vacua that were randomly chosen in the total space of $2^{45}$ vacua. However, the classification of flipped $SU(5)$ models in Ref. \cite{frs} compared to here was also accomplished using a JAVA code that scanned $10^{12}$ string vacua in the space of $2^{44}$ vacua. The reason for having 45 free parameters here compared to the 44 free parameters in Ref. \cite{frs} is due to the GGSO coefficient $C\binom{z_2}{z_2}$ being dependent on the string spectrum unlike before. This is a property emerging from modular invariance that constrains every GUT gauge group model different from its given basis vectors. Therefore, different flipped $SU(5)$ $SO(10)$ breaking basis vectors were implemented here, to find three-generation models that were surprisingly all projected out in Ref. \cite{frs} in exophobic vacua. Furthermore, from every generated model desired phenomenological criteria are analyzed. As in the previous classifications, this paper also considers viable phenomenological criteria such as finding three-generation models, GUT breaking heavy Higgs, electroweak breaking light Higgs, exophobic vacua and an anomaly-free gauge group as its primary aim. For this reason, the observable sector of a heterotic-string free-fermionic  flipped $SU(5)$ model is characterized by the following 15 integers,
$\left(n_1,n_{\overline{1}},n_{5s},n_{\overline{5}s},n_{10},n_{\overline{10}},n_g,n_{{10H}},
n_{\overline{5}v},n_{5v},n_{{5h}},n_{1e},n_{\overline{1}e},n_{5e},n_{\overline{5}e}\right)$:
\begin{align*}
	&n_{1}\,\,\,\,\,\,=\,\,\,\text{\# of } ({\textbf 1},{+\textstyle\frac{5}{2}}),\\
	&n_{\overline{1}}\,\,\,\,\,\,=\,\,\,\text{\# of } ({\textbf 1},{-\textstyle\frac{5}{2}}),\\
	&n_{5s}\,\,\,\,=\,\,\,\text{\# of } ({\textbf 5},{+\textstyle\frac{3}{2}}),\\
	&n_{\overline{5}s}\,\,\,\,=\,\,\,\text{\# of } ({\overline{\textbf 5}},{-\textstyle\frac{3}{2}}),\\
	&n_{10}\,\,\,\,=\,\,\,\text{\# of } ({\textbf {10}},{+\textstyle\frac{1}{2}}),\\
	&n_{\overline{10}}\,\,\,\,=\,\,\,\text{\# of } ({\overline{\textbf {10}}},{-\textstyle\frac{1}{2}}),\\
	&n_g\,\,\,\,\,\,=\,\,\,n_{10}-n_{\overline{10}}\,=\,\,\,n_{\overline{5}}-n_{5}=\,\,\,
	\text{\# of generations,}\\
	&n_{{10H}}=\,\,\,n_{10} + n_{\overline{10}}\,=\,\,\,\text{\# of nonchiral heavy Higgs pairs,}\\
	&n_{\overline{5}v}\,\,\,\,=\,\,\,\text{\# of } ({\overline{\textbf 5}},{+1}),\\
	&n_{5v}\,\,\,\,=\,\,\,\text{\# of } ({\textbf 5},{-1}),\\
	&n_{{5h}}\,\,\,\,=\,\,\,n_{5v} + n_{\overline{5}v}=\,\,\,\text{\# of nonchiral light Higgs pairs,}\\
	&n_{1e}\,\,\,\,\,=\,\,\,\text{\# of } ({\textbf 1 },
	{-\tfrac{ 5}{ 4}})\text{ \,\, (exotic),}\\
	&n_{\overline{1}e}\,\,\,\,\,=\,\,\,\text{\# of } ({{\textbf 1}},
	{+\tfrac{ 5}{ 4}})\text{ \,\, (exotic),}\\
	&n_{5e}\,\,\,\,\,=\,\,\,\text{\# of } ({\textbf 5},
	{ -\tfrac{1}{4}} )\ \text{ \,\,(exotic),}\\
	&n_{\overline{5}e}\,\,\,\,\,=\,\,\,\text{\# of } ({\overline{\textbf 5}},
	{+\tfrac{ 1}{ 4}}) \text{ \,\, (exotic).}
\end{align*}
These numbers above are also identical to the criteria used in Ref. \cite{frs} that were used for the classification and to extract viable data for low-energy physics at the string scale. A unique feature of the $\alpha$ projection is that it always projects the states $n_{\overline{1}}$, $n_{5s}$ and $n_{1}$, $n_{\overline{5}s}$ in or out together and therefore they are also given as pairs. This reduces the total number of integers determining viable vacua from 15 to 13. Additionally, the distinction is made with the flipped $SU(5)$ $\textbf5$ and $\overline{\textbf{5}}$ representations decomposing from the $SO(10)$ spinorial $\textbf{16}$ and vectorial $\textbf{10}$ representations, where the Standard Model up-type quark electroweak singlet and lepton doublet are denoted by $n_{5s}$ and $n_{\overline{5}s}$ respectively and the light electroweak Higgs doublets are denoted by the pair $n_{5v},n_{\overline{5}v}$. In order for the models to be semirealistic, further constraints are imposed on the integers given above; these include $n_{\overline{5}h} = n_{5h}$, $n_{1e} = n_{\overline{1}e}$ and
$n_{5e} = n_{\overline{5}e}$. This ensures that the flipped $SU(5)$ models are anomaly free. For the models to be phenomenologically viable the following are also imposed:
\begin{align*}
	&n_g=3~~~~~~~~~~~~\text{Three light chiral number of generations.}\\
	&n_{{10H}}\ge1~~~~~~~~~\text{At least one heavy Higgs pair to 
		break the $SU(5)\times U(1)$ symmetry.}\\
	&n_{{5}h}\ge1~~~~~~~~~~~\text{At least one pair of light minimal SM Higgs doublets.}\\
	&n_{1e}=n_{\overline{1}e} \ge 0~~~
	\text{Heavy mass can be generated for vector--like
		exotics.}\\
	&n_{5e} = n_{\overline{5}e} \ge 0~~~
	\text{Heavy mass can be generated for vector--like
		exotics.}\\
\end{align*}
This completes all the details necessary for the classification of the flipped $SU(5)$ models to be carried out and the methodology being applied. In the next section, the classification results are given and a discussion is carried out.

\subsection{\emph{Results}}

\begin{center}
	\begin{figure}[H]
		\includegraphics[width=.9\linewidth]{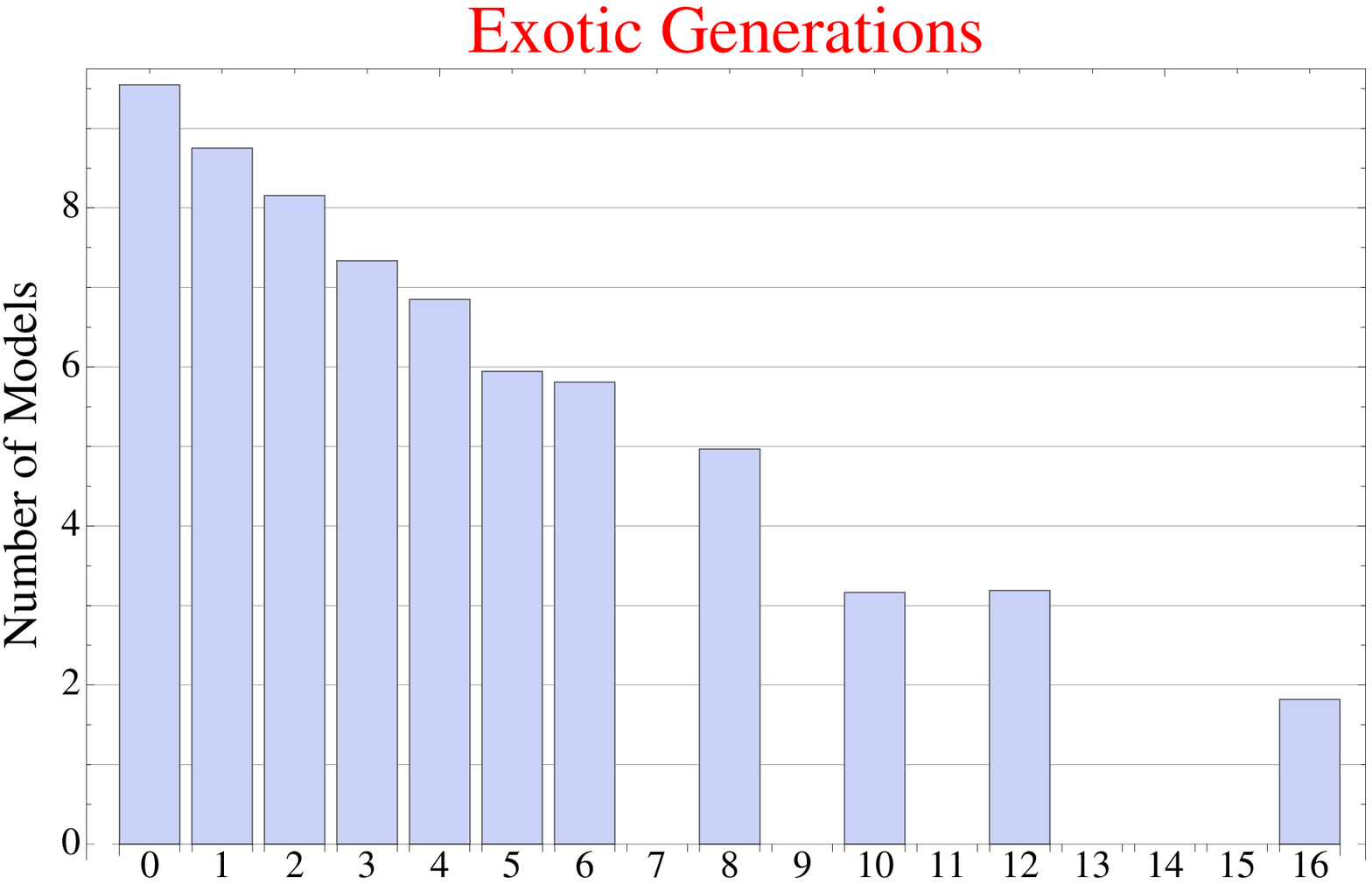}
		\caption{Logarithmic distribution of the number of exophilic flipped $SU(5)$ models against the number of chiral generations in a random sample of $10^{11}$ distinct configurations.}
	\end{figure}
\end{center}

\begin{center}
	\begin{figure}[H]
		\includegraphics[width=.9\linewidth]{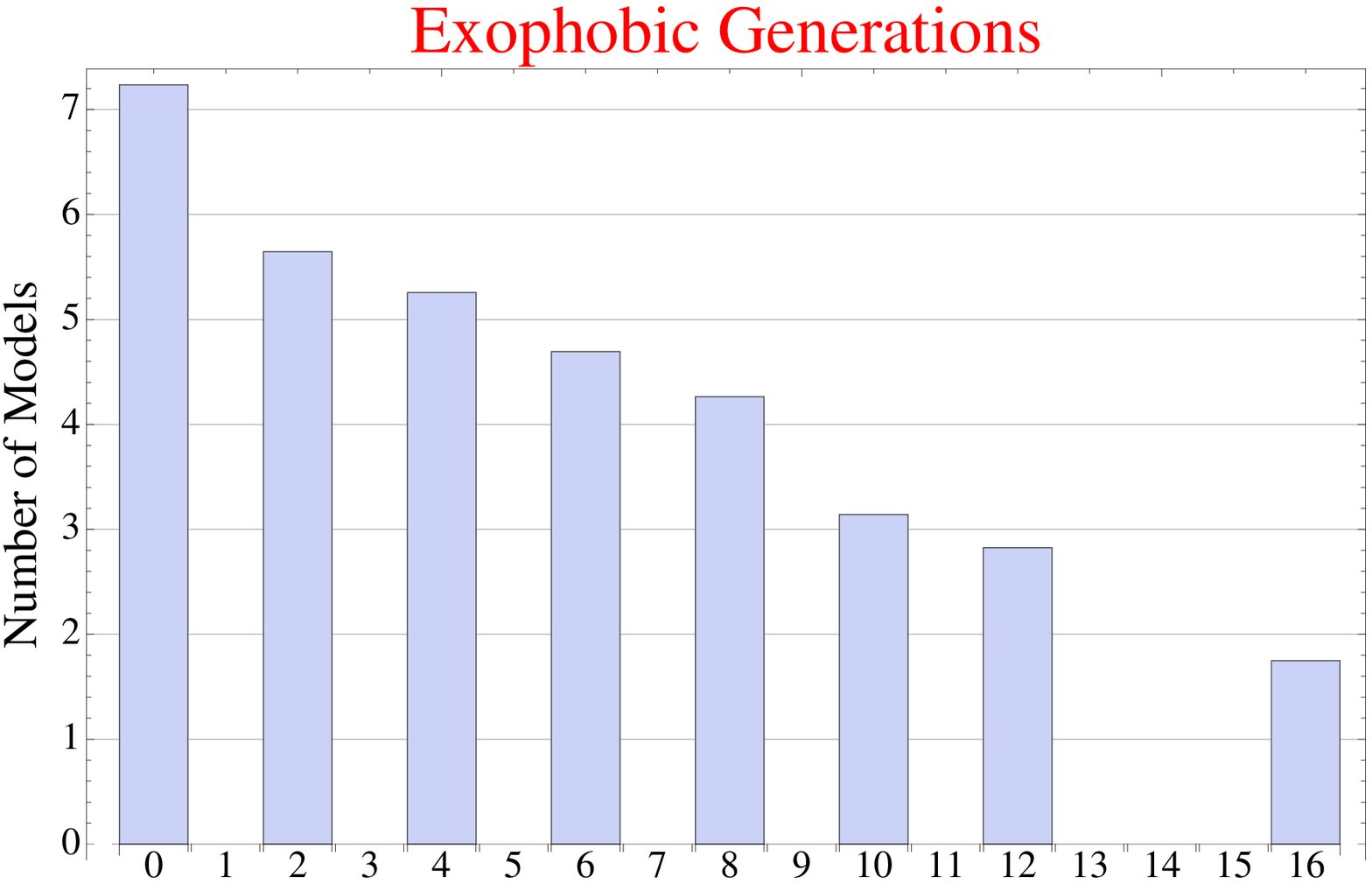}
		\caption{Logarithmic distribution of the number of exophobic flipped $SU(5)$ models against the number of chiral generations in a random sample of $10^{11}$ distinct configurations.}
	\end{figure}
\end{center}

\begin{center}
	\begin{figure}[H]
		\includegraphics[width=.9\linewidth]{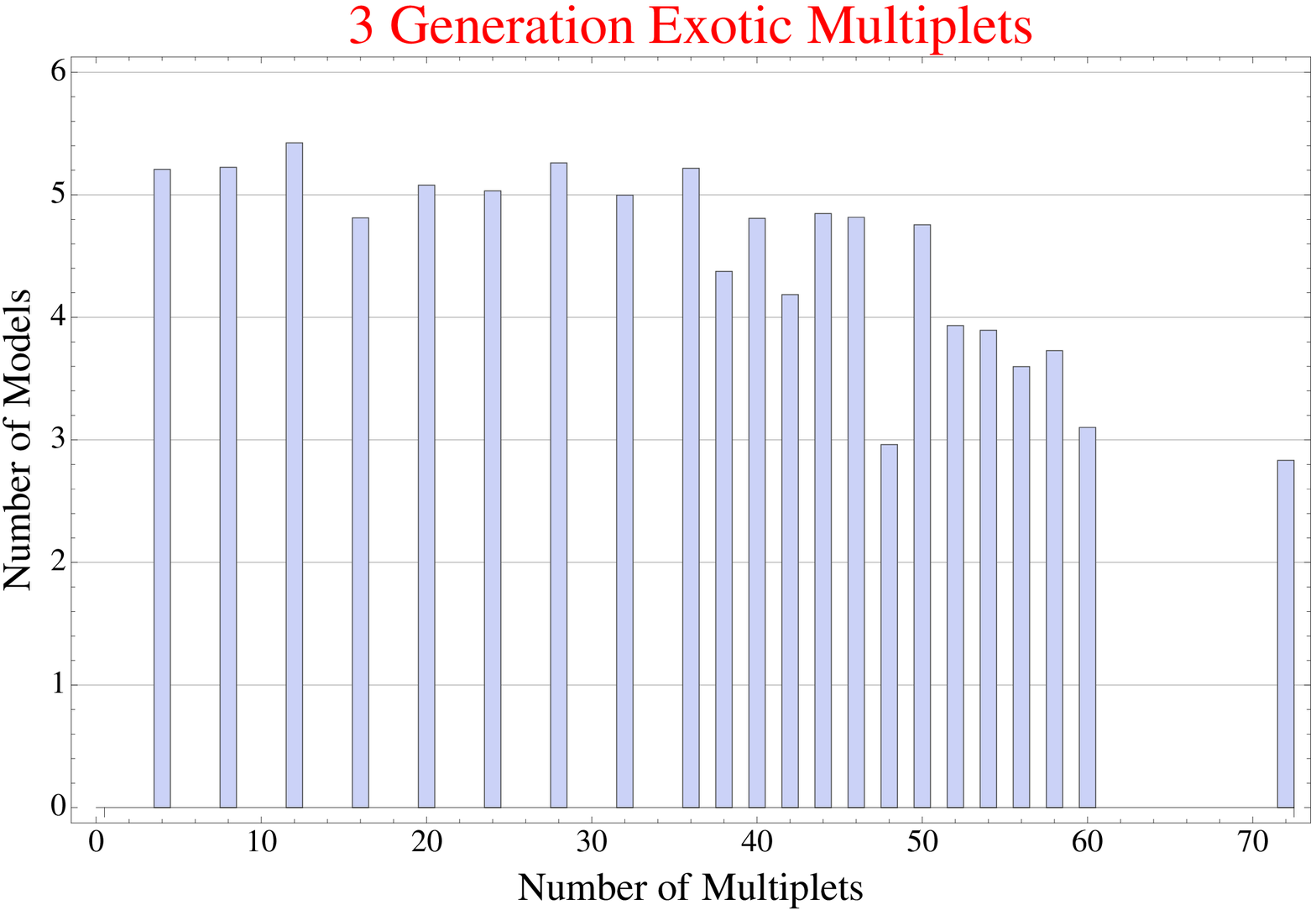}
		\caption{Logarithmic distribution of the number of three-chiral-generation flipped $SU(5)$ models against the number of exotic multiplets in a random sample of $10^{11}$ distinct configurations.}
	\end{figure}
\end{center}

In this section, the classification of the free-fermionic flipped $SU(5)$ string vacua is discussed and compared to the one in Ref. \cite{frs}; the classification was carried out by statistical sampling of $10^{11}$ models out of the $2^{45} \approx 3.52 \times 10^{13}$ vacua. In Ref. \cite{frs} the scan was carried out for $10^{12}$ models out of the $2^{42} \approx 1.76 \times 10^{13}$ vacua. The scan was accomplished here with a JAVA code similarly to Ref.  \cite{frs}, which essentially solves algebraic equations as discussed in section 3.4 and then checks for viable phenomenological data from the 13 integers discussed in the previous section. The JAVA routine was run using the nodes at the University of Liverpool, Department of Physics ULGQCD cluster and several servers in the University of Liverpool, Department of Mathematics. Results were gained in a week and are presented in this section with Figs. 1-3 and Table 1.

In Fig. 1, the logarithmic distribution of the number of exophilic flipped $SU(5)$ models against the number of chiral generations in a random sample of $10^{11}$ distinct configurations is displayed. The peak is seen at zero generations and as the number of generations increases the number of models decreases logarithmically. The figure also presents the absences of 7, 9, 11, 13, 14 and 15 generations and it should be noted that there are no models with more than 16 generations. In comparison to Fig. 1 in Ref. \cite{frs}, 16-generation models arise here and were excluded in Ref. \cite{frs}; this is due to the extra freedom given by the hidden gauge group enhancements. It is concluded here that no significant changes are observed with a variation of the flipped $SU(5)$ $SO(10)$ breaking basis vector. Additionally, the results here are in agreement with Refs. \cite{frs,PhDThesis,FKNR2004a,FKNR2004b,FKR2007a,FKR2007b,FKR2007c,Rizos2011,ACFKR2011a,ACFKR2011b,SU421,DiscretePaper}.

In Fig. 2, the logarithmic distribution of the number of exophobic flipped $SU(5)$ models against the number of chiral generations in a random sample of $10^{11}$ distinct configurations is displayed. The striking feature given in this figure is the repeat of the nonexistence of three chiral generations from Ref. \cite{frs}. The figure also presents the absences of odd generations, in addition to the exclusion of 14 and above 16 generations. In comparison to Fig. 2 in Ref. \cite{frs}, 16-generation models again arise here and were excluded in Ref. \cite{frs}, due to the extra freedom given by the hidden gauge group enhancements. It is concluded that no significant changes are observed with a variation of the flipped $SU(5)$ $SO(10)$ breaking basis vector and results are in agreement with Refs. \cite{frs,PhDThesis,FKNR2004a,FKNR2004b,FKR2007a,FKR2007b,FKR2007c,Rizos2011,ACFKR2011a,ACFKR2011b,SU421,DiscretePaper}. It should be emphasized that these results hold in the space of models that were explored here and that it does not necessarily indicate the absence of three-generation exophobic flipped $SU(5)$ models; however it might indicate the difficulty of finding them if they exist.

In Fig. 3, the logarithmic distribution of the number of three-chiral-generation flipped $SU(5)$ models against the number of exotic multiplets in a random sample of $10^{11}$ distinct configurations is displayed. Similarly to Ref. \cite{frs}, the figure shows that the minimal number of exotic multiplets is four; therefore as three chiral generations exist in exophilic flipped $SU(5)$ vacua and until experimental data can dictate otherwise, all three-chiral-generation flipped $SU(5)$ models will be imposed with a minimal number of exotic multiplets (which here is four).

\begin{table}[H]
	\begin{tabular}{|c|l|r|c|c|r|}
		\hline
		&Constraint & \parbox[c]{2.5cm}{\# Of Models In \\ Sample Scanned }& Probability
		&\parbox[c]{3cm}{ Estimated \# Of \\ Models In Class}\\
		\hline
		& Total Models & $100000000000$ & $1$ &$3.52\times 10^{13}$ \\ \hline
		(1)&{+ Anomoly-Free Models} & 8010089227 & $8.01\times 10^{-2}$ & $2.82\times
		10^{12}$ \\  \hline
		(2)&{+ No Enhancements} & 6590765377 & $6.59\times 10^{-2}$ & $2.32\times
		10^{12}$ \\  \hline
		(3)&{+ Number of Three-Generation Models} & 20929202 & $2.09 \times 10^{-4}$ & $7.36\times
		10^{9}$ \\  \hline
		(4)&{+ SM Light Higgs Breaking} & 20094915 & $2.09 \times 10^{-4}$ & $7.07\times
		10^{9}$ \\\hline
		(5)&{+ FSU5 Heavy Higgs Breaking} & 1677071 & $1.68 \times 10^{-5}$ & $5.90\times
		10^{8}$ \\\hline
		(6)&+ FSU5 Heavy Higgs Breaking \& & 1597702 & $1.60 \times 10^{-5}$ & $5.62\times
		10^8$ \\
		& SM Light Higgs Breaking&&&\\\hline
		(7)&{+ Minimal FSU5 Heavy Higgs Breaking} & 1550194 & $1.55 \times 10^{-5}$ & $5.45\times
		10^8$ \\\hline
		(8)&{+ Minimal SMLight Higgs Breaking} & 831508 & $8.32 \times 10^{-6}$ & $2.93\times
		10^8$ \\\hline
		(9)&+ Minimal FSU5 Heavy Higgs \&  & 811564 & $8.12 \times 10^{-6}$ & $2.86\times
		10^8$ \\
		&SM Light Higgs Breaking&&&\\\hline
		(10)&{+ Minimal Exotic States} & 90364 & $9.04 \times 10^{-7}$ & $3.18\times
		10^7$ \\\hline
	\end{tabular}
	\caption{Table consisting of total number of viable flipped $SU(5)$ models with respect to the phenomenological constraints imposed given by a sequence of increasing numbers. Details include: constraints imposed with respect to low-energy physics, the remaining number of models with the imposed constraints, the probability of the number of models occurring with the imposed constraints and the estimated number of models occurring in the space of $2^{45}$ vacua.}
\end{table}

In Table 1, the total number of viable flipped $SU(5)$ models with respect to the phenomenological constraints imposed given by a sequence of increasing numbers is displayed.
The column ``Constraint" shows the sequence of low-energy physics constraints imposed. The column ``\# Of Models In Sample Scanned" shows the remaining models in the scanned sample as each constraint is imposed. The column ``Probability" shows the probability of the occurrence of models with respect to the imposed constraints in the sample scanned. The column ``Estimated \# Of Models In Class" predicts the number of models in the entire space with such imposed constraints. The scan was carried out for a sample of $10^{11}$ vacua and the initial tabulation after imposing that only anomaly-free flipped $SU(5)$ models with no gauge group enhancement of the four-dimensional gauge symmetry, shows that only approximately $6.59\times 10^{-2}\%$ of the models remain. Next, imposing the existence of both the heavy and light Higgs states to break the flipped $SU(5)$ gauge symmetry to the Standard Model gauge group and the electroweak breaking respectively, leads to a further reduction to approximately $1.60 \times 10^{-5}\%$ of the models. Additional reduction is achieved by imposing minimal flipped $SU(5)$ heavy Higgs, SM light Higgs and minimal exotic states, which leaves $9.04 \times 10^{-7}\%$ of the models. These models are believed to be semirealistic models in nature; an example of such a model is given by the following matrix of one-loop phases for a minimal three-generation model:
\begin{equation} \label{BigMatrix} \ \ \bordermatrix{
		& S&e_1&e_2&e_3&e_4&e_5&e_6&b_1&b_2&z_1&z_2&\alpha\cr
		S  		& -1& -1& -1& -1& -1& -1& -1& -1& -1& -1& -1& -1\cr
		e_1		& -1& \,\,\,\,1& \,\,\,\,1& -1& -1& -1& \,\,\,\,1& \,\,\,\,1& \,\,\,\,1& -1& -1& -1\cr
		e_2		& -1& \,\,\,\,1& \,\,\,\,1& \,\,\,\,1& \,\,\,\,1& -1& \,\,\,\,1& -1& \,\,\,\,1& -1& -1& -1\cr
		e_3		& -1& -1& \,\,\,\,1& \,\,\,\,1& \,\,\,\,1& \,\,\,\,1& \,\,\,\,1& \,\,\,\,1& \,\,\,\,1& \,\,\,\,1& -1& \,\,\,\,1\cr
		e_4		& -1& -1& \,\,\,\,1& \,\,\,\,1& \,\,\,\,1& \,\,\,\,1& \,\,\,\,1& \,\,\,\,1& \,\,\,\,1& \,\,\,\,1& -1& -1\cr
		e_5		& -1& -1& -1& \,\,\,\,1& \,\,\,\,1& \,\,\,\,1& -1& \,\,\,\,1& -1& -1& -1& -1\cr
		e_6		& -1& -1& \,\,\,\,1& \,\,\,\,1& \,\,\,\,1& -1& \,\,\,\,1& -1& -1& \,\,\,\,1& -1& -1\cr
		b_1		& \,\,\,\,1& \,\,\,\,1& -1& \,\,\,\,1& \,\,\,\,1& \,\,\,\,1& -1& \,\,\,\,1& \,\,\,\,1& -1& -1& \,\,\,\,i\cr
		b_2		& \,\,\,\,1& \,\,\,\,1&\,\,\,\, 1& \,\,\,\,1& \,\,\,\,1& -1& -1& \,\,\,\,1& -1&\,\,\,\, 1&\,\,\,\, 1&\,\,\,\, i\cr
		z_1		& -1& -1& -1&\,\,\,\, 1&\,\,\,\, 1& -1& \,\,\,\,1& -1& \,\,\,\,1& -1& -1& -1\cr
		z_2		& -1& -1& -1& -1& -1& -1& -1& -1& \,\,\,\,1& -1&\,\,\,\, 1& -i\cr
		\alpha	& -1& -1& -1& \,\,\,\,1& -1& -1& -1&\,\,\,\, 1& \,\,\,\,1& \,\,\,\,1& \,\,\,\,1& -1\cr
	}\nonumber. 
\end{equation}
The attribute of this model is $n_g = 3$, $n_{\overline{5}s} = 3$,  $n_{5s} = 0$, $n_{10} = 4$, $n_{\overline{10}}=1$, $n_{10H} = 1$, $n_{5h} = 4$, $n_{1e} = 2$ and $n_{5e} = 0$. In conclusion,
unlike the classification of the free-fermionic Pati-Salam models \cite{ACFKR2011b}, which contained three-chiral-generation models free of massless exotic states, the free-fermionic flipped $SU(5)$ models both in Ref. \cite{frs} and in this paper show that no three-chiral-generation models exist and that they only exist with the inclusion of four exotic multiplets as given in the model above. Although this was a surprising feature, it is however in line with related searches \cite{GatoRivera2011}.

\section{\emph{Conclusion}}
The ${\mathbb{Z}}_2 \times {\mathbb{Z}}_2$ free-fermionic orbifolds in four dimensions that are given at special points in the moduli space, have demonstrated many semirealistic string models constructed to date \cite{RevampAEHN,SLMa, SLMb, SLMc, SLMd, SLMe, LRS2001a, LRS2001b, NonSusyPaper,ALR1990a, ALR1990b}. This has led to the development of the symmetric ${\mathbb{Z}}_2 \times {\mathbb{Z}}_2$ free-fermionic orbifold classification methodology given in Refs. \cite{frs,PhDThesis,FKNR2004b,FKR2007a,FKR2007b,FKR2007c,spinvecduala,spinvecdualb,spinvecdualc,spinvecduald,Rizos2011,ACFKR2011a,ACFKR2011b,CFR2011,SU6SU2,SU421}. All these models adapted the classification methodology by containing the six additional basis vectors $e_i$ for $i=1,...,6$ that consist of all the possible symmetric ${\mathbb{Z}}_2$ shifts which are in the internal compactified directions. Moreover the $e_i$ basis vectors allowed the writing of algebraic equations that could be computed by a computer code, which inevitably enabled the scanning of a large number of free-fermionic string vacua. The first classification in Ref. \cite{FKNR2004a} scanned the chiral $\bf{16}$ and $\overline{\bf{16}}$ spinorial $SO(10)$ GUT representations in order to show viable phenomenology. The Higgs states were then tested in Ref. \cite{FKR2007a} with the scanning of $SO(10)$ GUT $\bf{10}$ vectorial representations. As more developments were made the spinor-vector duality \cite{FKR2007b, FKR2007c} discrete symmetry was discovered using the $x$-map \cite{XMAP1993a,XMAP1993b}.
From these developments the classification methodology was then known to elegantly extract the full massless spectrum from a set of configurations, with the writing of the GGSO projections in algebraic forms to a computer code. As a result, the $SO(10)$ models \cite{Rizos2011} were scanned in detail and became a success, as it was shown to have an abundance of three-generation models. When the classification was done for a random sample of $10^{11}$ string vacua, discrete properties began to emerge. Here, it was observed that the odd generations above five vanished, whereas the even generations above 12 were incremented by four integers. 

The advent of the $SO(10)$ models saw the Pati-Salam models being investigated. The Pati-Salam models \cite{ACFKR2011b} contained identical discrete properties as the $SO(10)$ models, with the exclusion of all the 24 generations, since they were all projected out. In this case, as the spinorial $\bf{16}$ representation of the $SO(10)$ was broken to the $\bf{4}$ and $\overline{\bf{4}}$ representations transforming under the $SU(4)$ of the Pati-Salam, therefore, instead of one, two states were required to complete the family of generations. However, the Pati-Salam models also contained many three-generations that were exophobic. Furthermore, the flipped $SU(5)$ models were then explored both in Ref. \cite{frs} and here. Contrary to the Pati-Salam classification consisting of exophobic three-generation models, the flipped $SU(5)$ models contained only three-generation models with fractionally charged exotic states, in a random sample  of $10^{12}$ and $10^{11}$ string vacua scanned. An additional property of the flipped $SU(5)$ models, is that they were more constrained, as the generations followed a logarithmic distribution together with all the odd generations being projected out. In hopes of finding viable three-generation exophobic flipped $SU(5)$ string vacua, the next and final flipped $SU(5)$ breaking basis vector is under investigation; as discussed in section 2, this basis vector is given in the following form:
\begin{equation*}
	\alpha \, = \{ \overline{\eta}^{1,2,3} = \frac{1}{2}, \overline{\psi}^{1,...,5} = \frac{1}{2}, \overline{\phi}^{1,2} = \frac{1}{2}, \overline{\phi}^{5,6} = \frac{1}{2}, \overline{\phi}^{8} = 1\}.
\end{equation*}
Furthermore, in the event of not achieving the desired goal of finding viable three-generation exophobic flipped $SU(5)$ string vacua, the symmetric ${\mathbb{Z}}_2 \times {\mathbb{Z}}_2$ free-fermionic orbifolds will be reevaluated and a new set of $SO(10)$ breaking basis vectors will be constructed. In this new construction an additional hidden gauge group breaking basis will be added to the ones in Eq. (\ref{basis}), which is given by
\begin{equation*}
	z_3 \, = \{ \overline{\phi}^{1,2}, \overline{\phi}^{5,6}\}.
\end{equation*}
However, it should be noted that models consisting of these three hidden gauge group breaking basis vectors have not been studied before and are more complicated, and therefore results might take considerably more time to access. 

In this paper, the discrete properties that emerged in the landscape of the free-fermionic heterotic-string classifications were presented. Emphasis should be put on the fact that a JAVA routine was used to classify the flipped $SU(5)$ models and that no three-generation exophobic vacua were found as in Ref. \cite{frs}. Therefore, other possibilities are currently being explored. In the pursuit of finding such three-generation flipped $SU(5)$ models, two $SO(10)$ breaking basis vectors are also under investigation. These result in the $SU(4) \times SU(2) \times U(1)$, standard-like and left-right symmetric models. The classification of the $SU(4) \times SU(2) \times U(1)$ models was in fact explored already and it was revealed that these models were even more constrained than the flipped $SU(5)$ models. However, this was anticipated, due to the presence of two $SO(10)$ breaking basis vectors, which forbid complete generations. In actual fact, this was a rare occurrence in the free-fermionic classifications, as the second $SO(10)$ breaking basis vector was unique and the GGSO projection on the $\bf{16}$ of $SO(10)$ projected out all the right-handed particles. Thus, a whole family of generations was incomplete. The next stage will be the classification of the standard-like, left-right symmetric and the remaining flipped $SU(5)$ models in the free-fermionic construction; these are works in progress and are being considered extensively for future publications.

To conclude, the current status of the unification of gravity with the gauge interactions seems to be heavily motivated by string theory which continues to provide a viable consistent framework. Consequently, the three-generation string models need to be obtained for phenomenological purposes. Having said that, we are a long way away from pinning down a detailed example. Nevertheless, string theory provides a sea of such established semirealistic examples which are explored as toy models in the quest for a theory of everything.

\section{\emph{Acknowledgements}}
H.S. would like to thank Johar M. Ashfaque, Alon E. Faraggi and John Rizos for the many useful discussions. H. S. would also like to thank the String Phenomenology 2015 organisers hosted in Madrid (IFT-UAM/CSIC) for their hospitality.

\end{document}